# Electron-electron interactions and weak anti-localization in few-layer ZrTe$_5$ devices

Zhijian Xie,[1] Xinjian Wei,[1,2] Shimin Cao,[1,2] Yu Zhang,[2] Shili Yan,[2] G. D. Gu,[3] Qiang Li,[3,4] Jian-Hao Chen[1,2,5*]

[1]*International Center for Quantum Materials, School of Physics, Peking University, Beijing 100871, China*
[2]*Beijing Academy of Quantum Information Sciences, Beijing 100193, China*
[3]*Condensed Matter Physics & Materials Science Division, Brookhaven National Laboratory, Upton, New York 11973-5000, USA*
[4]*Department of Physics and Astronomy, Stony Brook University, Stony Brook, NY 11794-3800, USA*
[5]*Key Laboratory for the Physics and Chemistry of Nanodevices, Peking University, Beijing 100871, China*

*E-mail: *Jian-Hao Chen (chenjianhao@pku.edu.cn)*

## ABSTRACT

Much effort has been devoted to the electronic properties of relatively thick ZrTe$_5$ crystals, focusing on their three-dimensional topological effects. Thin ZrTe$_5$ crystals, on the other hand, were much less explored experimentally. Here we present detailed magnetotransport studies of few-layer ZrTe$_5$ devices, in which electron-electron interactions and weak anti-localization are observed. The coexistence of the two effects manifests themselves in corroborating evidence presented in the temperature and magnetic field dependence of the resistance. Notably, the temperature-dependent phase coherence length extracted from weak anti-localization agrees with strong electron-electron scattering in the sample. Meanwhile, universal conductance fluctuations have temperature and gate voltage dependence that is similar to that of the phase coherence length. Lastly, all the transport properties in thin ZrTe$_5$ crystals show strong two-dimensional characteristics. Our results provide new insight into the highly intricate properties of topological material ZrTe$_5$.

## I.   INTRODUCTION

The transition metal pentatelluride ZrTe$_5$ has attracted tremendous interests due to its unusual topological nature and fascinating properties[1,2]. Angle-resolved photoelectron spectroscopy (ARPES) and transport experiments reveal that ZrTe$_5$ is a material in proximity to the topological phase transition between weak and strong topological insulator[3-8]. At the boundary of weak and strong topological insulator, ZrTe$_5$ exhibits properties of Dirac semimetal with linear dispersion[9-11] and could become a Weyl semimetal if time reversal symmetry is broken[12]. Many transport studies have demonstrated its unusual topological nature: chiral magnetic effect[13], anomalous Hall effect[12], planar Hall effect[14] and three-dimensional quantum Hall effect[1]. Interestingly, the band structure of ZrTe$_5$ can be tuned by strain, temperature and thickness[2,7,15,16]. As a layered compound, ZrTe$_5$ is stacked by van der Waals interaction, which makes it easily to be exfoliated. Theoretically, monolayer ZrTe$_5$ is a good candidate for large gap quantum spin Hall insulator[17], which is promising for spintronics applications due to the potentially dissipationless edge transport. Currently, while there are abundant studies on bulk ZrTe$_5$ crystal properties, there is much less investigation in its two-dimensional (2D) limit. In this article we present magnetotransport studies of few layer ZrTe$_5$ devices at low temperature. Electron-electron (e-e) interaction and weak anti-localization (WAL) are found to coexist in thin ZrTe$_5$ devices at the 2D limit and provide intricately corroborating experimental evidence. Specifically, logarithmic increase in the longitudinal resistivity is observed with a slope consists of contributions from e-e interactions and WAL[18]; at higher temperature, a $T^2$ increase in the resistivity arised from the Fermi liquid behavior of the sample[19,20]; a clear sign of 2D WAL is observed at the weak field magnetoresistance at low temperatures and provides phase coherence length with a power-law dependence in temperature which is consistent with the existence of e-e interactions; finally, the root-mean-square (rms) of universal conductance fluctuation of the sample in a sweeping magnetic field shows a power-law dependence with temperature, with an exponent value supporting e-e interaction in the sample. At the meantime, strong 2D characteristics are observed in these samples. In additional to 2D WAL, we found that the magnetoresistance (MR) signal and UCF rms are both proportional only to the perpendicular component of the magnetic field.

## II. EXPERIMENTAL METHODS

The ZrTe$_5$ crystals are grown by the flux growth method using Te as flux[13], and thin ZrTe$_5$ flakes are mechanically exfoliated on silicon substrate with 285nm oxide. Standard e-beam lithography was used to pattern electrodes followed by an e-beam evaporation of Cr (5 nm) and Au (60 nm). The electrical measurements were performed in a physical property measurement system (PPMS) with standard lock-in technique. Special attention has been paid to protect the thin flakes from exposing to ambient conditions. All the sample preparation process and device fabrication process were done in vacuum, inert atmosphere, or with the sample capped with a protection layer. The protection layer consists of a bilayer of 200 nm PMMA and 200 nm MMA.

## III. RESULTS AND DISCUSSION

Fig. 1a shows the temperature-dependent resistance of a typical few-layer ZrTe$_5$ device (labeled as device 1). It can been seen that the resistance decreases with decreasing temperature monotonically from room temperature down to low temperature (~10K), contrary to the temperature-dependent resistivity of bulk and thick ZrTe$_5$ samples, where a highly characteristic resistivity peak is widely observed[2]. The temperature $T_0$ at which the maximum resistivity shows up depends on crystal doping, sample thickness and magnetic field[13,15,16]. In a specific crystal under zero magnetic field, $T_0$ is associated with sample thickness. In general, $T_0$ moves to higher temperature in thinner samples[15,16]. In our thin samples (thickness 21.0 nm, see inset in Fig. 1a for the section height profile obtained by Atomic Force Microscopy), the resistivity peak is missing below room temperature, which is consistent with previous reports in thin ZrTe$_5$ flakes[15,16]. We also observed that at temperatures below 10K (region marked by red square box in Fig. 1a), the resistivity increases logarithmically with decreasing temperature, which is consistent with e-e interaction in disordered system[21-23]. The correction to classical Drude conductance (here we use $\sigma(T = 10\text{K})$ as a good estimation) from e-e interaction and WAL can

be theoretically described by Equation (1)[18,24]:

$$\Delta\sigma(T) = C\frac{e^2}{\pi h}\ln T \qquad (1)$$

where $e$ is elementary charge, $h$ is the Plank constant, the coefficient $C = -\alpha_1 p + 1 - 3/4 \cdot F$. Equation (1) is a consequence of competition between corrections from WAL and e-e interactions to the low temperature conductance. WAL, which we shall discuss later, gives rise to the term $\alpha_1 p$ while e-e interaction contributes the term $1 - 3/4 \cdot F$ in $C$[18,22]. The Hikami prefactor $\alpha_1$ is supposed to be 1/2, $p$ can be extracted from the power law temperature dependence of the phase coherence length $L_\varphi \propto T^{-p/2}$ and $F$ is the screening factor. Our WAL analysis shows that $p = 1$. As can be seen in Fig.1b, the conductance of device 1 at low temperatures can be well described by Equation 1. From the fitting, the parameter $C$ is determined to be 0.82, which gives rise to $1/2 - 3/4 \cdot F = 0.82$, in contradictory with the fact that $F$ should lie between 0 and 1. The determination of sample length/width ratio in calculating the sheet conductance may cause as much as 23% of uncertainty in $C$ (see the Supplemental Material for details[25]), which, by itself, could not explain our observation. Meanwhile, such discrepancy is also observed in Cu-doped $Bi_2Se_3$ film for a $C$ factor of 1.4, which is interpreted as excess contribution from bulk state together with the surface state[18,21,22,26]. At elevated temperature (50K~200K), resistivity follows the behavior of $\rho = \rho_0 + AT^2$. This can be understood in the framework of Fermi liquid theory[19,20,27]. In Fermi liquid with e-e scattering as the dominant scattering mechanism, resistivity would exhibit quadratic dependence on temperature[19,20,27]. In particular, it is predicted that the temperature dependent scattering rate should be $\frac{1}{\tau} = \frac{A}{\hbar}\frac{(k_B T)^2}{\varepsilon_F}$, where $A$ is a dimensionless factor, $k_B$ is Boltzmann constant, $\hbar$ the reduced Plank constant, $\varepsilon_F$ is the Fermi energy[27]. Using the Drude model $\sigma = ne^2\tau/m^*$ and temperature dependence of carrier density $n$ (inset of Fig. 1c) we can extract the temperature dependent scattering rate as shown in Fig. 1c, which agrees quite well with the theory. Here we adopt the effective mass $m^* = 0.07 m_e$[15], $m_e$ is the mass of free electron. Taking the Fermi energy $\varepsilon_F$ to be ~35.4meV from previous report[10], the dimensionless factor $A$ would be 1.2, which falls reasonably in the theoretically predicted value (between unity to 100[27]). The effect of electron-phonon scattering is not considered since the scattering rate from the electron-phonon scattering is primarily $1/\tau \sim T^3$ theoretically[27]. There are some reports of $1/\tau \sim T^2$ in high-resistance granular samples[28] (different from our single crystal $ZrTe_5$ sample), which occur at liquid-helium temperature. Phonon in $ZrTe_5$ at the temperature range of 50K - 260K might have contributed to a temperature dependent scattering rate of $1/\tau \sim T^\delta$ ($\delta \geq 3$), but it is dominated by a $1/\tau \sim T^2$ term that does not come from phonon scattering. The next available scattering mechanism at this temperature range is e-e interaction.

We then measured the magnetotransport properties of thin $ZrTe_5$ at 2K with perpendicular magnetic field. Positive *MR* is observed, as shown in Fig.1d, similar to many other topological materials[29-32]. The magnitude of the *MR* is much smaller in few-layer $ZrTe_5$ than in bulk crystals[13], similar to the observation in ultra-thin $WTe_2$[33] as compares to bulk $WTe_2$[29]. Hall curves in thin flakes do not show non-linear behavior, different from that of thick flakes or bulk materials[10,34], indicating strong dependence of the band structure and/or Fermi surface to sample thickness. From the Hall data we extract the mobility and carrier density of the sample to be 1119 cm$^2$/Vs and 1.70×10$^{13}$/cm$^2$.

At small magnetic field, a sharp cusp in MR can be seen (Fig. 1d) which has strong temperature dependence (see Fig. 2a for $\delta\sigma$ vs. $T$). This is a characteristic of the WAL effect, e.g., in a system where time reversal symmetry is preserved, the electron wave function of a back scattering path interferes constructively (weak localization, short for WL) or destructively (weak anti-localization, short for WAL) with its time reversal counterpart, which enhances or suppresses the probability of electron localization. When the time reversal symmetry is broken by a small magnetic field, the quantum interference would be destroyed and therefore leads to positive magnetoconductance (in the case of WL) or negative magnetoconductance (in the case of WAL) [35]. WL is generally presented in two dimensional electronic system, and WAL appears if the system has strong spin-orbit interaction or a nontrivial Berry phase[35]. In three dimension, WL/WAL exists as well, with a different functional form[36].

The temperature dependence of WAL is presented in Fig. 2a. WAL quickly vanishes with increasing temperature. As the temperature raises to 30K, the magnitude of WAL is rather small. To gain insight into the dephasing mechanism of the WAL in ZrTe$_5$, we fit the magnetoconductance by the Hikami-Larkin-Nagaoka (HLN) model which formulates the quantum correction of the magnetoconductance in two dimensions [37]:

$$\Delta\sigma = -\alpha_2 \frac{e^2}{\pi h}\left(\Psi\left(\frac{1}{2}+\frac{B_\varphi}{B}\right) - \ln\frac{B_\varphi}{B}\right) \qquad (2)$$

where $\Psi$ is the digamma function, $B_\varphi = \hbar/(4eL_\varphi^2)$ is the characteristic magnetic field related to the phase coherence of the electrons, $L_\varphi$ is the phase coherence length, $B$ represents the magnetic field and Hikami prefactor $\alpha_2$ is an empirical fitting parameter, which equals to 1/2 and -1 for single channel transport in the WAL and WL regimes, respectively. The experimental data in Fig. 2a can be satisfactorily fitted by Equation 2, giving strong evidence that the electronic system is two dimensional (more evidence of the dimensionality of the electronic system of thin ZrTe$_5$ will be show later in the text).

From the HLN fitting one can obtain the phase coherence length $L_\varphi$. We plot the temperature dependence of $L_\varphi$ in Fig. 2b. $L_\varphi$ tends to saturate to around 130 nm at the lowest temperature then decrease with a power law behavior $L_\varphi \sim T^{-0.51\pm0.02}$ with increasing temperature to 52 nm at 30K. In general, the dephasing behavior of $L_\varphi$ is $L_\varphi \sim T^{-\beta}$. The exponent $\beta$ varies in different scattering mechanisms. In the 2D case, $\beta = 1$ for electron-phonon scattering and $\beta = 1/2$ for electron-electron scattering[24]. Thus, the experimentally measured $\beta = -0.51 \pm 0.02$ in our device is consistent with the theoretical expectation from e-e scattering in thin ZrTe$_5$. We note that the fitting results of the prefactor in Equation 2, $\alpha_2$, is around 1 (see inset of Fig. 2b), this may ascribe to two-channel transport. $\alpha_2$ obtained from a thinner device (thickness 11.7 nm, see the Supplemental Material for details[25]) is around 0.6, this may be due to the coupling of different channels in thinner samples. $\alpha_1$ in Eqn.(1) and $\alpha_2$ in Eqn.(2) characterize WAL correction on temperature and on magnetic field, respectively. Interestingly, we find that the extracted value of $\alpha_1 \sim 0.5$ from Eqn. (1) (see the Supplemental Material for details[25]) is smaller than $\alpha_2 \sim 1$ from Eqn. (2), similar to the results from previous reports in Bi$_2$Se$_3$ thin films[22,23], which warrants further theoretical and experimental investigations.

To further verify the 2D nature of the WAL in our device, the angular dependence of the magnetoresistance with tilted magnetic field is investigated. Fig. 2c depicts the magnetoconductance with perpendicular component of magnetic field, i.e. $B\cos\theta$ at different tilt angles, $\theta$ is the angle between the direction of magnetic field and the direction perpendicular to the device plane. It can be seen in the main panel of Fig. 2c that for small magnetic field, all curves at different angles collapse to a single trace. This indicates that the observed WAL is the result of the electron motion in the 2D plane. For *MR* in larger magnetic field as shown in the inset of Fig. 2c, the $\delta\sigma$ vs. $B\cos\theta$ curves have some small mismatch between each other, indicating additional effects of the in-plane magnetic field. At the meantime, the Hall coefficient $R_H$ with titled magnetic field shows very nice dependence to $1/\cos\theta$ as plotted in Fig. 2d, which provides another solid evidence of the 2D nature of the electronic states in thin $ZrTe_5$.

We then investigate the WAL dependence on carrier density tuned by the silicon back gate. Due to the 2D nature of the sample, albeit relatively high hole concentration in the sample, the resistance and carrier density can still be tuned via the back gate (inset of Fig. 3b), resulting in significant modification to the WAL signal. Some representative traces of WAL at different back gate voltages $V_{bg}$ in a second device we measured (device 2-1, thickness 11.7nm, see the Supplemental Material for details[25]) at 2K are shown in Fig. 3a. The magnetoconductance correction of WAL can be tuned by nearly threefold, from $0.2e^2/h$ at $V_{bg} = 100V$ to $0.6e^2/h$ at $V_{bg} = -100V$. The corresponding phase coherence length changes from 250 nm to 95 nm. From Fig. 3b we can see that the phase coherence length increases monotonically with increasing carrier concentration. This is observed in other 2D materials and topological insulators[24,38-41]. Such behavior can be qualitatively explained by the Altshuler-Aronov-Khmelnitsky theory which relates the sheet conductance $\sigma$ and inelastic scattering time $\tau_\varphi$ as: $\frac{\hbar}{\tau_\varphi} = \frac{k_B T}{\sigma/\sigma_q}\ln(\sigma/\sigma_q)$, where $\sigma_q$ is the conductance quantum $\frac{e^2}{h}$[24,42]. Combined with $L_\varphi = \sqrt{D\tau_\varphi}$, where $D = \frac{\hbar}{4m^*}\frac{\sigma}{\sigma_q}$ is diffusive coefficient, we can get:

$$L_\varphi = \hbar\sigma/\sigma_q[\ln(\sigma/\sigma_q)\,4m^*k_B T]^{-1/2} \qquad (3)$$

As higher carrier density leads to higher conductivity, naturally the phase coherence length is enhanced (also see the plot of the experimentally obtained $L_\varphi$ vs. $(\sigma/\sigma_q)/\sqrt{\ln\sigma/\sigma_q}$ in Supplemental Material[25]).

Universal conductance fluctuation (UCF) is another manifestation of quantum mechanical interference of electronic wave function in mesoscopic systems. We observed magnetoconductance fluctuations in thin $ZrTe_5$ devices at large magnetic fields and low temperatures which can be attributed to UCF. The fluctuations are repeatable with multiple sweeps of the magnetic field (see the Supplemental Material[25]) and diminish with increasing temperature. The conductance fluctuations $\delta G$ (after subtracting a smooth background) versus magnetic field of the thin $ZrTe_5$ device 2-2 at $V_{bg} = 0$ are presented in Fig. 4a at different temperatures. The amplitude of conductance fluctuation can be described by its root mean square (rms), defined as $\text{rms}(G) = \sqrt{<\delta G(B)*\delta G(B)>}$, where $<\cdots>$ means ensemble average. Fig. 4b shows the temperature dependence of rms(*G*) at $V_{bg} = 0$. Rms*(G)* is theoretically predicted to be $\sqrt{\frac{L_\varphi^2}{LW}}\frac{e^2}{h}$ at zero temperature, where $L$ and $W$ is the length and width of sample, respectively[43], from which we expect the rms(*G*) to be $\sim 0.08e^2/h$, close to the experimental result. Since rms(*G*) is theoretically predicted to be proportional to $L_\varphi$ for 2D systems with sample dimension $L \gg L_\varphi$ [44,45], one can expect a completely identical behavior of the dependence of rms(*G*) and $L_\varphi$ on external parameters such as temperature and gate voltage, which is exactly what we observed. Fig. 4b shows the temperature dependent rms(*G*) (left panel) and $L_\varphi$ (right panel) obtained from device 2-2 at $V_{bg} = 0$. $L_\varphi$ is extracted from WAL at small magnetic field, while rms(*G*) is measured from UCF from high field magnetoconductance. It can be seen that the functional behavior of rms(*G*) and $L_\varphi$ are indeed identical, indicating that both WAL and UCF is associated with the same phase coherence length. Consistently, rms(*G*) has very similar gate voltage dependence to that of $L_\varphi$ (Fig. 4c & 4d). By rotating the magnetic field, we identify that the UCF is also two dimensional, similar to WAL. Fig. 4e depicts the conductance fluctuation in different angles. As the magnetic field rotates towards the in-plane direction, the position of representative peak or valley marked by circles in Fig. 4e in the conductance fluctuations moves to higher magnetic field. The magnetic field corresponding to peak or valley in $\delta G$ can be well fitted by $1/\cos\theta$ (see Fig. 4f for three peak/valley positions, marked as B, C & D in Fig. 4e, versus $1/\cos\theta$), confirming that the fluctuation is only associated with perpendicular component of the magnetic field.

Lastly, we discuss the origin of the observed WAL. WAL is usually observed in systems with strong spin orbit coupling. Besides, WAL occurs in 3D topological insulator thin layers (also have strong spin orbit coupling), which is believed to arise from the nontrivial $\pi$ Berry phase of the topological surface state[23,46]. Considering the lack of smoking gun parameters in the WAL behavior alone to distinguish between topologically trivial and non-trivial electronic systems, we shall not make a deterministic statement in this paper; instead, we will provide some intriguing facts that might help the on-going debate as for whether ZrTe$_5$ is a strong or weak topological insulator[3-8,10]. There are three hints we found in this experiment that are leaning towards the possibility of a strong TI scenario (albeit NOT a proof of its existence): First, WAL is absent in thick devices while electron-electron interaction is still observed (see the Supplemental Material[25]). Along the line of a (hypothesized) STI, surface state should indeed contribute a smaller fraction of the total signal for thicker sample for the WAL effect, while e-e interaction is less affected as it has contribution from the whole sample. Second, the Hikami prefactor $\alpha_2$ in Eqn. (2) is found to be ~1 in a 21.0nm sample while its value is ~0.6 in a 11.7nm sample, which is in the same direction of stronger coupling of the top and bottom surfaces states for the thinner sample. Third, the phase coherence length tuned by carrier density predicted by Eqn. (3) is larger than experimentally observed (see the Supplemental Material[25]), indicating that only a portion of the carrier density induced by gate voltage participates in the WAL transport. Given the above observations, it is still hard to distinguish the exact topological classification of ZrTe$_5$, and more experiments are needed to fully clarify this issue.

## IV. CONCLUSION

In summary, we have observed the coexistence of weak anti-localization and electron-electron interaction in thin ZrTe$_5$ devices with multiple transport evidence. From the HLN model, the phase coherence length is 95-250 nm at low temperatures and exhibits a power law dependence on temperature. The main dephasing mechanism in our devices is determined to be electron-electron interactions. Universal conductance fluctuations are also observed and its behavior corroborates the conclusion from the WAL analysis. Thin ZrTe$_5$ samples are also found to be highly two-dimensional in terms of their electronic properties. Our results provide new insight into the highly intricate properties of the topological material ZrTe$_5$.

## V. ACKNOWLEDGEMENT

The authors thank X. C. Xie, S. Q. Shen and L. Y. Zhang for helpful discussion. This project has been supported by the National Basic Research Program of China (Grant Nos. 2019YFA0308402，2018YFA0305604), the National Natural Science Foundation of China (NSFC Grant Nos. 11934001，11774010，11921005), Beijing Municipal Natural Science Foundation (Grant No. JQ20002). The work at Brookhaven National Laboratory was supported by the U. S. Department of Energy (DOE), Office of Basic Energy Sciences, Division of Materials Sciences and Engineering, under Contract No. DE-SC0012704.

†Corresponding author: chenjianhao@pku.edu.cn

## REFERENCE

1. F. Tang, Y. Ren, P. Wang, R. Zhong, J. Schneeloch, S. A. Yang, K. Yang, P. A. Lee, G. Gu, Z. Qiao and L. Zhang. Three-dimensional quantum Hall effect and metal-insulator transition in $ZrTe_5$. *Nature* **569**, 537-541 (2019).

2. Y. Zhang, C. Wang, L. Yu, G. Liu, A. Liang, J. Huang, S. Nie, X. Sun, Y. Zhang, B. Shen, J. Liu, H. Weng, L. Zhao, G. Chen, X. Jia, C. Hu, Y. Ding, W. Zhao, Q. Gao, C. Li, S. He, L. Zhao, F. Zhang, S. Zhang, F. Yang, Z. Wang, Q. Peng, X. Dai, Z. Fang, Z. Xu, C. Chen and X. J. Zhou. Electronic evidence of temperature-induced Lifshitz transition and topological nature in $ZrTe_5$. *Nat. Commun.* **8**, 15512 (2017).

3. X. B. Li, W. K. Huang, Y. Y. Lv, K. W. Zhang, C. L. Yang, B. B. Zhang, Y. B. Chen, S. H. Yao, J. Zhou, M. H. Lu, L. Sheng, S. C. Li, J. F. Jia, Q. K. Xue, Y. F. Chen and D. Y. Xing. Experimental observation of topological edge states at the surface step edge of the topological insulator $ZrTe_5$. *Phys. Rev. Lett.* **116**, 176803 (2016).

4. Y.-Y. Lv, B.-B. Zhang, X. Li, K.-W. Zhang, X.-B. Li, S.-H. Yao, Y. B. Chen, J. Zhou, S.-T. Zhang, M.-H. Lu, S.-C. Li and Y.-F. Chen. Shubnikov–de Haas oscillations in bulk $ZrTe_5$ single crystals: Evidence for a weak topological insulator. *Phys. Rev. B* **97**, 115137 (2018).

5. H. Xiong, J. A. Sobota, S. L. Yang, H. Soifer, A. Gauthier, M. H. Lu, Y. Y. Lv, S. H. Yao, D. Lu, M. Hashimoto, P. S. Kirchmann, Y. F. Chen and Z. X. Shen. Three-dimensional nature of the band structure of $ZrTe_5$ measured by high-momentum-resolution photoemission spectroscopy. *Phys. Rev. B* **95**, 195119 (2017).

6. R. Wu, J. Z. Ma, S. M. Nie, L. X. Zhao, X. Huang, J. X. Yin, B. B. Fu, P. Richard, G. F. Chen, Z. Fang, X. Dai, H. M. Weng, T. Qian, H. Ding and S. H. Pan. Evidence for topological edge states in a large energy gap near the step edges on the surface of $ZrTe_5$. *Phys. Rev. X* **6**, 021017 (2016).

7. J. Mutch, W.-C. Chen, P. Went, T. Qian, I. Z. Wilson, A. Andreev, C.-C. Chen and J.-H. Chu. Evidence for a strain-tuned topological phase transition in $ZrTe_5$. *Sci. Adv.* **5**, eaav9771 (2019).

8. G. Manzoni, L. Gragnaniello, G. Autes, T. Kuhn, A. Sterzi, F. Cilento, M. Zacchigna, V. Enenkel, I. Vobornik, L. Barba, F. Bisti, P. Bugnon, A. Magrez, V. N. Strocov, H. Berger, O. V. Yazyev, M. Fonin, F. Parmigiani and A. Crepaldi. Evidence for a strong topological insulator phase in $ZrTe_5$. *Phys. Rev. Lett.* **117**, 237601 (2016).

9. R. Y. Chen, S. J. Zhang, J. A. Schneeloch, C. Zhang, Q. Li, G. D. Gu and N. L. Wang. Optical spectroscopy study of the three-dimensional Dirac semimetal $ZrTe_5$. *Phys. Rev. B* **92**, 075107 (2015).

10. G. L. Zheng, J. W. Lu, X. D. Zhu, W. Ning, Y. Y. Han, H. W. Zhang, J. L. Zhang, C. Y. Xi, J. Y. Yang, H. F. Du, K. Yang, Y. H. Zhang and M. L. Tian. Transport evidence for the three-dimensional Dirac semimetal phase in $ZrTe_5$. *Phys. Rev. B* **93**, 115414 (2016).

11. A. Pariari and P. Mandal. Coexistence of topological Dirac fermions on the surface and three-dimensional Dirac cone state in the bulk of $ZrTe_5$ single crystal. *Sci. Rep.* **7**, 40327 (2017).

12. T. Liang, J. J. Lin, Q. Gibson, S. Kushwaha, M. H. Liu, W. D. Wang, H. Y. Xiong, J. A. Sobota, M. Hashimoto, P. S. Kirchmann, Z. X. Shen, R. J. Cava and N. P. Ong. Anomalous Hall effect in $ZrTe_5$. *Nat. Phys.* **14**, 451-455 (2018).

13. Q. Li, D. E. Kharzeev, C. Zhang, Y. Huang, I. Pletikosic, A. V. Fedorov, R. D. Zhong, J. A. Schneeloch, G. D. Gu and T. Valla. Chiral magnetic effect in $ZrTe_5$. *Nat. Phys.* **12**, 550-554 (2016).

14. P. Li, C. H. Zhang, J. W. Zhang, Y. Wen and X. X. Zhang. Giant planar Hall effect in the Dirac semimetal $ZrTe_{5-\delta}$. *Phys. Rev. B* **98**, 121108(R) (2018).

15. J. Niu, J. Wang, Z. He, C. Zhang, X. Li, T. Cai, X. Ma, S. Jia, D. Yu and X. Wu. Electrical transport in nanothick $ZrTe_5$ sheets: From three to two dimensions. *Phys. Rev. B* **95**, 035420 (2017).

16. J. W. Lu, G. L. Zheng, X. D. Zhu, W. Ning, H. W. Zhang, J. Y. Yang, H. F. Du, K. Yang, H. Z. Lu, Y. H. Zhang and M. L. Tian. Thickness-tuned transition of band topology in $ZrTe_5$ nanosheets. *Phys. Rev. B* **95**, 125135 (2017).

17. H. M. Weng, X. Dai and Z. Fang. Transition-metal pentatelluride $ZrTe_5$ and $HfTe_5$: A paradigm for large-gap quantum spin Hall insulators. *Phys. Rev. X* **4**, 011002 (2014).

18. H. Z. Lu and S. Q. Shen. Finite-temperature conductivity and magnetoconductivity of topological insulators. *Phys. Rev. Lett.* **112**, 146601 (2014).

19. D. Van Der Marel, J. L. M. Van Mechelen and I. I. Mazin. Common Fermi-liquid origin of $T^2$ resistivity and superconductivity in n-type $SrTiO_3$. *Phys. Rev. B* **84**, 205111 (2011).

20. X. Lin, B. Fauqué and K. Behnia. Scalable $T^2$ resistivity in a small single-component Fermi surface. *Science* **349**, 945-948 (2015).

21. N. P. Breznay, H. Volker, A. Palevski, R. Mazzarello, A. Kapitulnik and M. Wuttig. Weak antilocalization and disorder-enhanced electron interactions in annealed films of the phase-change compound $GeSb_2Te_4$. *Phys. Rev. B* **86**, 205302 (2012).

22. Y. Takagaki, B. Jenichen, U. Jahn, M. Ramsteiner and K.-J. Friedland. Weak antilocalization and electron-electron interaction effects in Cu-doped $Bi_2Se_3$ films. *Phys. Rev. B* **85**, 115314 (2012).

23. J. Chen, X. Y. He, K. H. Wu, Z. Q. Ji, L. Lu, J. R. Shi, J. H. Smet and Y. Q. Li. Tunable surface conductivity in $Bi_2Se_3$ revealed in diffusive electron transport. *Phys. Rev. B* **83**, 241304(R) (2011).

24. Y. Shi, N. Gillgren, T. Espiritu, S. Tran, J. Yang, K. Watanabe, T. Taniguchi and C. N. Lau. Weak localization and electron–electron interactions in few layer black phosphorus devices. *2D Mater.* **3**, 034003 (2016).

25. See Supplemental Material at [url] for the estimation of uncertainty in length/width ratio of device 1, atomic force microscopy and optical image of device 2, weak anti-localization of device 2-2, the logarithmic temperature dependence of conductance with different magnetic field, relation between $L_\varphi \sim (\sigma/\sigma_q)/\sqrt{(\ln(\sigma/\sigma_q))}$, universal conductance fluctuations in device 2-2, absence of weak anti-localization in thick devices, EEI effect on temperature dependent resistance in thick sample, magnetoresistance and Hall resistance of device 1 at different temperatures, gate dependence of Hall resistance via back gate in device 1, universal magnetoconductance fluctuations vs. $B\cos\theta$.

26   T. Ando, A. B. Fowler and F. Stern. Electronic properties of two-dimensional systems. *Rev. Mod. Phys.* **54**, 437-672 (1982).

27   N. W. Ashcroft and M. N. David. Solid state physics. 346-348 (Thomas Learning, 1976).

28   J J Lin and J. P. Bird. Recent experimental studies of electron dephasing in metal and semiconductor mesoscopic structures. *J. Phys.: Condens. Matter* **14**, R501–R596 (2002).

29   M. N. Ali, J. Xiong, S. Flynn, J. Tao, Q. D. Gibson, L. M. Schoop, T. Liang, N. Haldolaarachchige, M. Hirschberger, N. P. Ong and R. J. Cava. Large, non-saturating magnetoresistance in WTe$_2$. *Nature* **514**, 205-208 (2014).

30   T. Liang, Q. Gibson, M. N. Ali, M. Liu, R. J. Cava and N. P. Ong. Ultrahigh mobility and giant magnetoresistance in the Dirac semimetal Cd$_3$As$_2$. *Nat. Mater.* **14**, 280-284 (2015).

31   X. Wang, Y. Du, S. Dou and C. Zhang. Room temperature giant and linear magnetoresistance in topological insulator Bi$_2$Te$_3$ nanosheets. *Phys. Rev. Lett.* **108**, 266806 (2012).

32   S. Cao, W. Ma, G. Zhai, Z. Xie, X. Gao, Y. Zhao, X. Ma, L. Tong, S. Jia and J. H. Chen. Anisotropic Raman spectrum and transport properties of AuTe$_2$Br flakes. *J. Phys.: Condens. Matter* **32**, 12LT01 (2020).

33   X. Liu, Z. Zhang, C. Cai, S. Tian, S. Kushwaha, H. Lu, T. Taniguchi, K. Watanabe, R. J. Cava, S. Jia and J.-H. Chen. Gate tunable magneto-resistance of ultra-thin WTe$_2$ devices. *2D Mater.* **4**, 021018 (2017).

34   F. D. Tang, P. P. Wang, P. Wang, Y. Gan, L. Wang, W. Zhang and L. Y. Zhang. Multi-carrier transport in ZrTe$_5$ film. *Chinese Physics B* **27**, 087307 (2018).

35   F. V. Tikhonenko, A. A. Kozikov, A. K. Savchenko and R. V. Gorbachev. Transition between electron localization and antilocalization in graphene. *Phys. Rev. Lett.* **103**, 226801 (2009).

36   B. Fu, H. W. Wang and S. Q. Shen. Quantum interference theory of magnetoresistance in Dirac materials. *Phys. Rev. Lett.* **122**, 246601 (2019).

37   S. Hikami, A. I. Larkin and Y. Nagaoka. Spin-orbit interaction and magnetoresistance in the two dimensional random system. *Prog. Theor. Phys.* **63**, 707-710 (1980).

38   J. Chen, H. J. Qin, F. Yang, J. Liu, T. Guan, F. M. Qu, G. H. Zhang, J. R. Shi, X. C. Xie, C. L. Yang, K. H. Wu, Y. Q. Li and L. Lu. Gate-voltage control of chemical potential and weak antilocalization in Bi$_2$Se$_3$. *Phys. Rev. Lett.* **105**, 176602 (2010).

39   K. Gopinadhan, Y. J. Shin, I. Yudhistira, J. Niu and H. Yang. Giant magnetoresistance in single-layer graphene flakes with a gate-voltage-tunable weak antilocalization. *Phys. Rev. B* **88**, 195429 (2013).

40   S.-P. Chiu and J.-J. Lin. Weak antilocalization in topological insulator Bi$_2$Te$_3$ microflakes. *Phys. Rev. B* **87**, 035122 (2013).

41   J. Zeng, S.-J. Liang, A. Gao, Y. Wang, C. Pan, C. Wu, E. Liu, L. Zhang, T. Cao, X. Liu, Y. Fu, Y. Wang, K. Watanabe, T. Taniguchi, H. Lu and F. Miao. Gate-tunable weak antilocalization in a few-layer InSe. *Phys. Rev. B* **98**, 125414 (2018).

42   B. L. Altshuleri, A. G. Aronovf and D. E. Khmelnitsky. Effects of electron-electron collisions with small energy transfers on quantum localisation. *J. Phys. C: Solid State Phys.* **15**, 7367-7386 (1982).

43  S. W. Stanwyck, P. Gallagher, J. R. Williams and D. Goldhaber-Gordon. Universal conductance fluctuations in electrolyte-gated $SrTiO_3$ nanostructures. *Appl. Phys. Lett.* **103**, 213504 (2013).

44  P. A. Lee, A. D. Stone and H. Fukuyama. Universal conductance fluctuations in metals: Effects of finite temperature, interactions, and magnetic field. *Phys. Rev. B Condens Matter* **35**, 1039-1070 (1987).

45  J. Lee, J. Park, J.-H. Lee, J. S. Kim and H.-J. Lee. Gate-tuned differentiation of surface-conducting states in $Bi_{1.5}Sb_{0.5}Te_{1.7}Se_{1.3}$ topological-insulator thin crystals. *Phys. Rev. B* **86**, 245321 (2012).

46  H.-T. He, G. Wang, T. Zhang, I.-K. Sou, G. K. Lwong, J.-N. Wang, H.-Z. Lu, S.-Q. Shen and F.-C. Zhang. Impurity effect on weak antilocalization in the topological insulator $Bi_2Te_3$. *Phys. Rev. Lett.* **106**, 166805 (2011).

**Figures and captions**

Figure 1

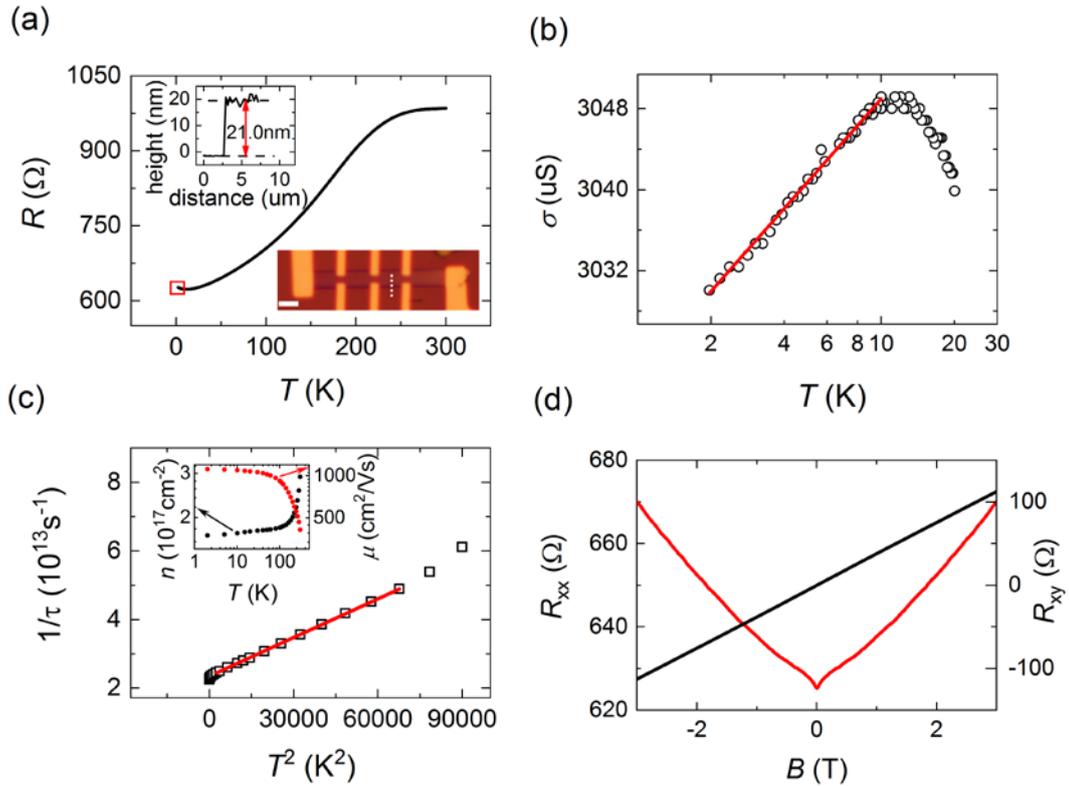

Figure 1. (a) Temperature dependent resistance of ZrTe$_5$ device 1. The upturn of the resistance at low temperature is marked by a small red box and is converted to conductance as shown in (b). The optical micrograph of the device is shown in the lower inset and the AFM section height data is shown in the upper inset. The scale bar in the lower inset is 5 um, and the white dashed line marks the position where the AFM section height data is acquired. (b) The conductance of device 1 as a function of temperature in semi-log scale. The red line is the linear fit to Equation 1. (c) Scattering rate $1/\tau$ as a function of $T^2$. The red line is a linear fit to the data at the temperature range of 50 K to 260 K. The inset depicts the carrier density (left axis, black dots) and mobility (right axis, red dots) at different temperatures. (d) Longitudinal (left axis, red curve) and Hall (right axis, black curve) resistance at $T = 2$K under perpendicular magnetic field.

Figure 2

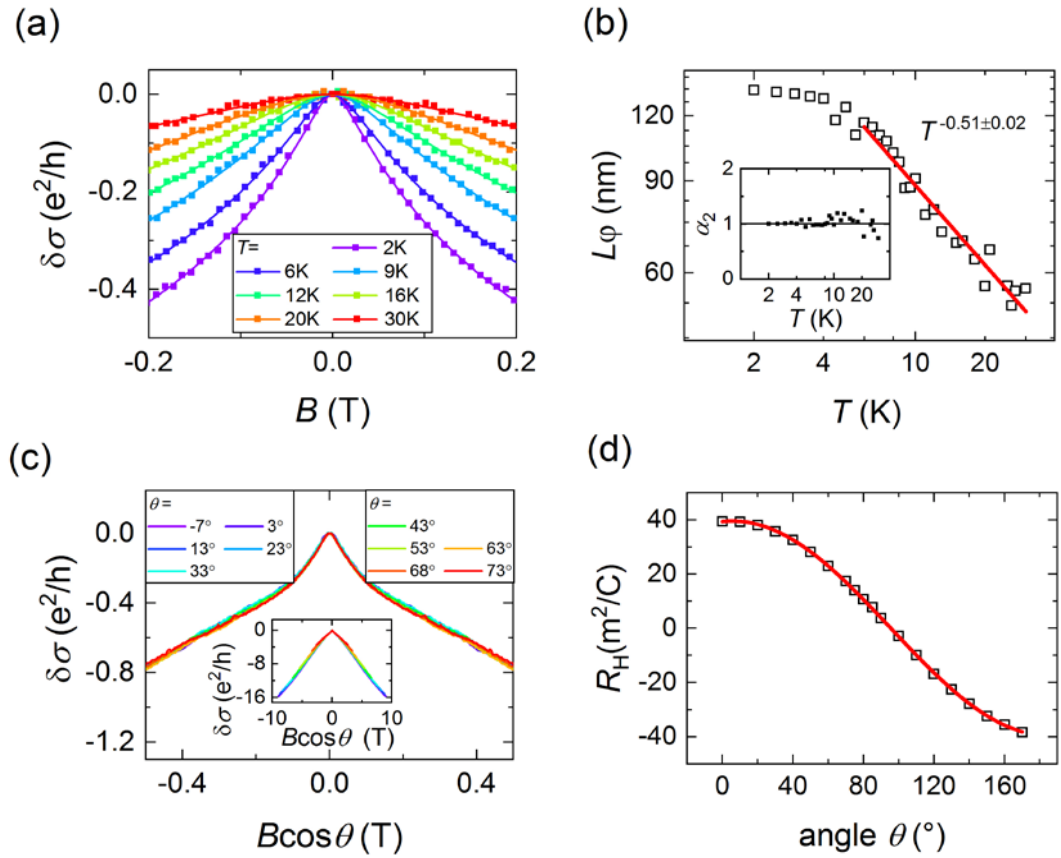

Figure 2. Weak anti-localization data of ZrTe$_5$ device 1. (a) Magnetoconductance $\delta\sigma$ at small magnetic field in the unit of $e^2/h$ at temperatures between 2K and 30K. Solid lines are fits to the experimental data using Equation (2). (b) The extracted phase coherence length from the fitting of figure 2(a). Inset: the fitting parameter $\alpha_2$ at different temperatures. (c) Magnetoconductance versus perpendicular component of the magnetic field with the sample titled at different angles at 2K. $\theta$ is the angle between the direction of the total magnetic field and the perpendicular axis of the sample plane. (d) The angular dependence of Hall coefficient $R_H$ and fit to $1/\cos\theta$. The angle is shifted by $7°$ for simplicity.

Figure 3

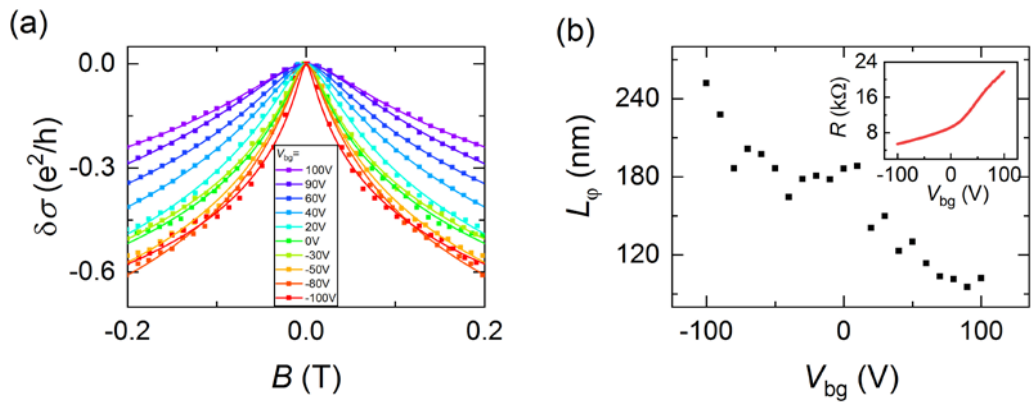

Figure 3. Gate dependence of WAL in ZrTe$_5$ device 2-1. (a) Representative traces of magnetoconductance with different back gate voltages at 2K. Solid lines are fits to the experimental data using Equation (2). (b) The extracted phase coherence length at various back gate voltages. Inset: gate dependence of resistance at $B = 0$.

Figure 4

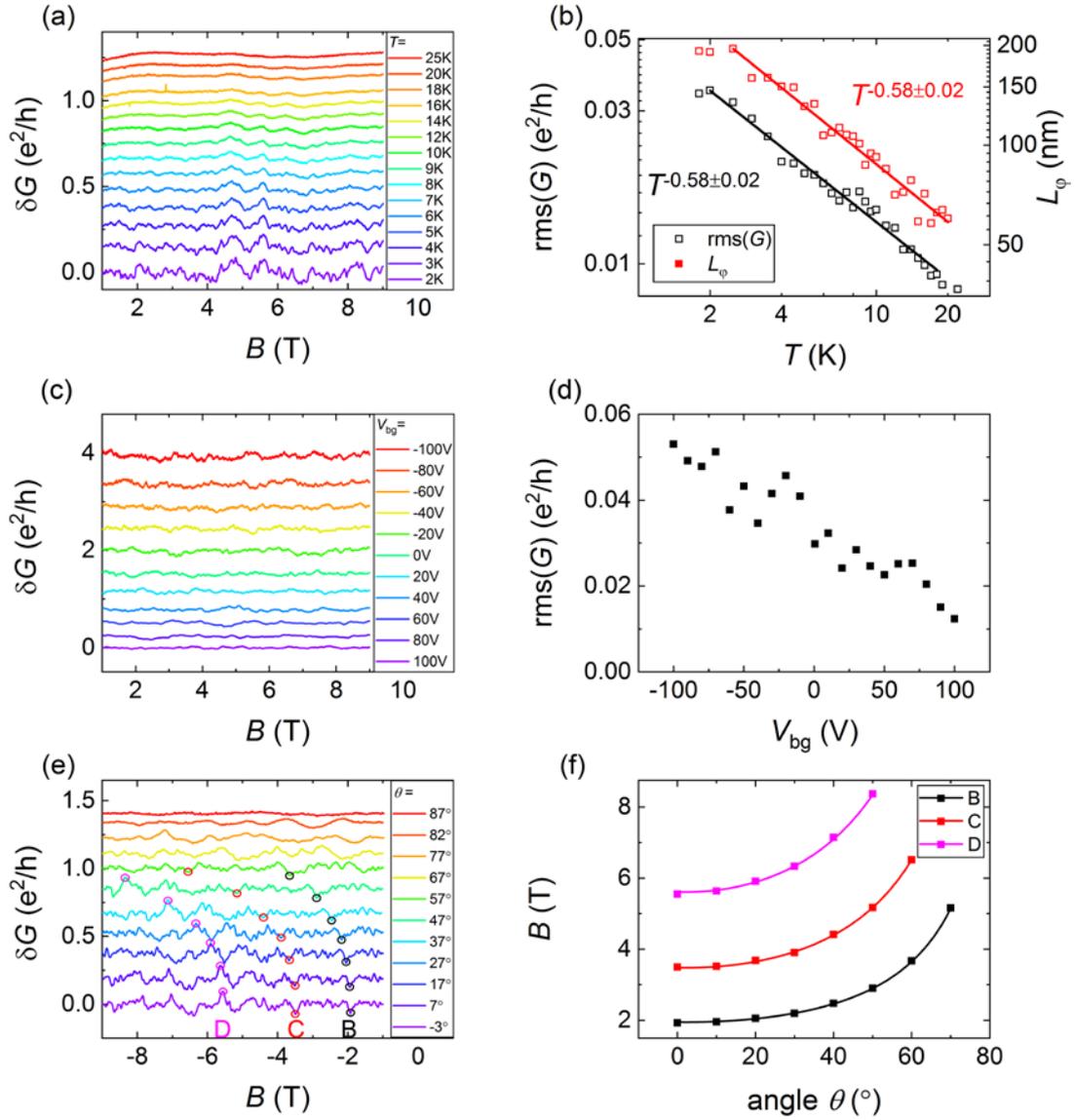

Figure 4. Universal conductance fluctuations in thin ZrTe$_5$. (a) Magnetoconductance fluctuation $\delta G$ in ZrTe$_5$ device 2-2 (after subtracting a smooth background, in units of $e^2/h$) versus magnetic field at different temperatures. Curves are shifted for clarity. (b) Root mean square of the magnetoconductance fluctuation in (a) as a function of temperature (left axis, black open squares); phase coherence length extracted from WAL of the same device (right axis, red solid squares). (c) Magneto-conductance fluctuation $\delta G$ at different back gate voltages at $T$ = 2K in device 2-1. (d) Gate dependence of rms($G$) from (c). (e) Magnetoconductance fluctuations at tilted magnetic field. Marked circles are several representative peaks or valleys of fluctuation. (f) The magnetic field corresponding to circles in (e) and their fits to $1/\cos\theta$ (solid lines). All the curves in figure 4f are shifted by 3° for simplicity.